\documentclass{article}

\usepackage{amssymb}

\usepackage{epsfig}

\usepackage{lscape}

\usepackage{graphicx,color}

\usepackage{graphicx,psfrag,amsmath}

\usepackage{yfonts}

\begin{document}

\def\beq#1\eeq{\begin{equation}#1\end{equation}}
\def\beql#1#2\eeql{\begin{equation}\label{#1}#2\end{equation}}

\def\bea#1\eea{\begin{eqnarray}#1\end{eqnarray}}
\def\beal#1#2\eeal{\begin{eqnarray}\label{#1}#2\end{eqnarray}}

\newcommand{\Z}{{\mathbb Z}}
\newcommand{\N}{{\mathbb N}}
\newcommand{\C}{{\mathbb C}}
\newcommand{\Cs}{{\mathbb C}^{*}}
\newcommand{\R}{{\mathbb R}}
\newcommand{\intT}{\int_{[-\pi,\pi]^2}dt_1dt_2}
\newcommand{\cC}{{\mathcal C}}
\newcommand{\cI}{{\mathcal I}}
\newcommand{\cN}{{\mathcal N}}
\newcommand{\cE}{{\mathcal E}}
\newcommand{\cA}{{\mathcal A}}
\newcommand{\xdT}{\dot{{\bf x}}^T}
\newcommand{\bDe}{{\bf \Delta}}

\def\ket#1{\left| #1\right\rangle }
\def\bra#1{\left\langle #1\right| }
\def\braket#1#2{\left\langle #1\vphantom{#2}
  \right. \kern-2.5pt\left| #2\vphantom{#1}\right\rangle }
\newcommand{\gme}[3]{\bra{#1}#3\ket{#2}}
\newcommand{\ome}[2]{\gme{#1}{#2}{\mathcal{O}}}
\newcommand{\spr}[2]{\braket{#1}{#2}}
\newcommand{\eq}[1]{Eq\,\ref{#1}}
\newcommand{\xp}[1]{e^{#1}}

\newcommand{\tend}[1]{$10^{#1}$}
\newcommand{\tennd}[1]{10^{#1}}
\newcommand{\nrd}[2]{${#1}\times{10^{#2}}$}
\newcommand{\nrnd}[2]{{#1}\times{10^{#2}}}

\def\limfunc#1{\mathop{\rm #1}}
\def\Tr{\limfunc{Tr}}

\def\dr{detector }
\def\drs{detectors }
\def\drsn{detectors}
\def\drn{detector}
\def\dtn{detection }
\def\dtnn{detection}

\def\pho{photon }
\def\phon{photon}
\def\phos{photons }
\def\phosn{photons}
\def\mmt{measurement }
\def\an{amplitude}
\def\a{amplitude }
\def\co{coherence }
\def\con{coherence}

\def\st{state }
\def\stn{state}
\def\sts{states }
\def\stsn{states}

\def\cow{collapse of the wavefunction }
\def\de{decoherence }
\def\den{decoherence}
\def\dm{density matrix }
\def\dmn{density matrix}

\newcommand{\mop}{\cal O }
\newcommand{\dt}{{d\over dt}}
\def\qm{quantum mechanics }
\def\qms{quantum mechanics }
\def\qml{quantum mechanical }

\def\qmn{quantum mechanics}
\def\mmtn{measurement}
\def\pow{preparation of the wavefunction }

\def\me{ L.~Stodolsky }
\def\T{temperature }
\def\Tn{temperature}
\def\t{time }
\def\tn{time}
\def\wfs{wavefunctions }
\def\wf{wavefunction }
\def\wfn{wavefunction} 
\def\wfsn{wavefunctions}
\def\wvp{wavepacket }
\def\pa{probability amplitude } 
\def\sy{system } 
\def\sys{systems }
\def\syn{system} 
\def\sysn{systems} 
\def\ha{hamiltonian }
\def\han{hamiltonian}
\def\rh{$\rho$ }
\def\rhn{$\rho$}
\def\op{$\cal O$ }
\def\opn{$\cal O$}
\def\yy{energy }
\def\yyn{energy}
\def\yys{energies }
\def\yysn{energies}
\def\pz{$\bf P$ }
\def\pzn{$\bf P$}
\def\pl{particle }
\def\pls{particles }
\def\pln{particle}
\def\plsn{particles}

\def\plz{polarization  }
\def\plzs{polarizations }
\def\plzn{polarization}
\def\plzsn{polarizations}

\def\sctg{scattering }
\def\sctgn{scattering}
\def\sctgs{scatterings }
\def\sctgsn{scatterings}

\def\prob{probability }
\def\probn{probability}

\def\om{\omega} 

\def\hf{\tfrac{1}{2}}
\def\hft{\tiny \frac{1}{2}}

\def\zz{neutrino }
\def\zzn{neutrino}
\def\zzs{neutrinos }
\def\zzsn{neutrinos}

\def\zn{neutron }
\def\znn{neutron}
\def\zns{neutrons }
\def\znsn{neutrons}

\def\hf{\tfrac{1}{2}}

\def\csss{cross-section }
\def\csssn{cross-section}
\def\xrn{x-ray nucleide }
\def\xrnn{x-ray nucleide}
\def\xr{x-ray }
\def\xrs{x-rays }
\def\wv{wavelength }
\def\wvn{wavelength}

\def\intf{interferometry }
\def\intfn{interferometry}
\def\intrf{interferometer }
\def\intrfn{interferometer}

\def\ran{radionuclide }
\def\rann{radionuclide}
\def\rans{radionuclides }

\def\dkm{dark matter }
\def\dkmn{dark matter}

\def\gvtl{gravitational }
\def\gvtln{gravitational}

\def\bh{black hole }
\def\bhn{black hole}
\def\bhs{black holes }
\def\bhsn{black holes}

\def\pbh{primordial black hole }
\def\pbhn{primordial black hole}
\def\pbhs{primordial black holes }
\def\pbhsn{primordial black holes}

\def\dph{$\delta \phi$ }
\def\dphn{$\delta \phi$}
\def\dphg{$\delta \phi_G$}
\def\dphgn{$\delta \phi_G$}

\def\gr{general relativity }
\def\grn{general relativity}

\def\mua{$\mu \rm arcsec$ }
\def\muan{$\mu arcsec$}
\def\rsl{resolution }
\def\rsln{resolution}

\def\lbi{long baseline interferometry }
\def\lbin{long baseline interferometry}

\def\imed{intervening medium }
\def\imedn{intervening medium}

\def\dg{\delta \phi_G}

\def\cd{column density }
\def\cdn{column density}

\title{Limits on Primordial Black Holes  from  M87}

\author{ 
Joseph Silk\\
Institut d'Astrophysique de Paris, UMR7095:CNRS \\
 UPMC-Sorbonne University, F-75014, Paris, France\\
\\
Leo Stodolsky\\
Max-Planck-Institut f\"ur Physik
(Werner-Heisenberg-Institut)\\
F\"ohringer Ring 6, 80805 M\"unchen, Germany}

\maketitle

\begin{abstract}  
Primordial black holes in the solar mass range
 are a possibly significant component of dark matter.
  We show how an argument relating the deflection of
 light by such black holes in the density spike  likely
 to exist around the  M87 supermassive black hole,
combined with the high resolution  observations of the EHT Collaboration,
can    lead to  strong limits
 on the primordial black hole mass fraction
 in an astrophysically relevant mass range. The results 
depend on the  model assumed for the dark matter spike and suggest the
interest of further understanding of such spikes as well as  further
high resolution observations on supermassive black holes. 
\end{abstract}

\section{Introduction}

Ascertaining the nature of dark matter remains one of
 the outstanding problems in cosmology. 
In recent times there have been active discussions of
 primordial black holes (PBH's)  as a possible
 dark matter candidate or as a component thereof.  
 These discussions are motivated both by the standard
  Lambda Cold Dark Matter (LCDM) model of cosmology,
 which can  plausibly generates PBHs in abundance, 
  and by the absence of any need
 to postulate new physics in order to establish
 their existence \cite{green} .

There have been numerous studies  that constrain the
 mass range of primordial black holes as a significant
  contributor to dark matter.
These include  gravitational microlensing \cite{machos}
 and ultrafaint dwarf galaxy heating \cite{dwarfs}. However none of these limits are conclusive, either because of the effects of PBH clustering  or because of uncertainties in  dwarf galaxy dynamical modelling \cite{models}. Indeed, it has equally been argued that the core/cusp transition in dwarfs may be due to PBH heating of cold dark matter  \cite{boldrini}. Here we consider a new limit that enables us to set 
 independent constraints on the fraction of asteroid to subsolar or even
 solar mass PBHs that can currently be a significant \dkm  (DM) contributor.

\section{Limits on \rsl due to \bhs}
If there is a high density of \bhs in some region of space the \gvtl fields
will be `lumpy' at short distances, inducing random deflections of light
rays. 
 These deflections  lead to a limitation on the
angular \rsl possible in observations, and if a high \rsl
is exhibited in some observation, there
 is then  a limit on the presence of \bhs along the flight
path.
 In ref \cite{sra} this effect was 
considered  for rays originating from the cosmic microwave background (CMB)
 and  in ref \cite{if} a related,  potentially
more sensitive  method,  applicable even when there are no small angle
features to observe, was suggested in terms of the  loss of coherence
between the arms of an interferometer,

However, for the application  presented here, the first method, in
terms of the angular \rsln, is the more favorable.
 The relevant formula for
the limit \cite{sra}  on the possible  angular \rsl $\delta \alpha_{lim}$ 
for rays passing through a \cd of \dkm $\Sigma $ is  
\beql{lim}
(\delta \alpha_{lim})^2\approx \nrnd{5}{-2} (M/M_{\odot})\, \Sigma_{ly} 
=\nrnd{5}{-3} (M/M_{\odot})\Sigma_{pc}\,,
\eeql
where the  \dkm is composed of 
\bhs of mass $M$. The angle $\delta \alpha_{lim}$ is
in \mua  and  
 $\Sigma$ on the right is in terms of
 %solar masses per $\rm ly^2$ as in \cite{sra}   or 
  solar masses per $\rm pc^2$, as we will use in the following. 
% as we will use here.

\eq{lim} is based on the reasoning that if 
 a \pho has a high probability,
approaching one, of being scattered  by a certain angle or more
during its flight, then an angular \rsl  of this amount is not possible.
The probability of such a \sctg is found from the product of the \cd with the \csss
corresponding to the formula  $\delta \alpha= 2 r_s/b$
($r_s$=Schwarzschild radius $=2GM,$
 $b$ is  impact parameter of the \pho, $c=1$ units),
 as in the bending of light by the sun.

A situation combining  very high \rsl observations with a suspected  high density
of \dkm arises for the Super Massive Black Hole (SMBH) in the galaxy M87.
The EHT collaboration has studied thia object  with a  \rsl of 
10 \mua \cite{eht}. At the same time it is likely \cite{Gondolo} 
that a  SMBH  is surrounded by a ``spike'' of \dkmn.
In the present note, we would like  to consider
the  implications  of \eq{lim} in this situation.

\section{Models around SMBHs:  M87}

To estimate $\Sigma$, a knowledge of
 the spike density profile is necessary. This profile is
  generated by adiabatic growth of the SMBH
 at early epochs. Its slope however  depends on the growth history.
 
We consider two alternative density profiles. One is 
 $\rho \propto r^{-7/3}$ for adiabatic growth of the SMBH by gas 
 accretion or stellar tidal disruption.
The ambient cold dark matter responds adiabatically and
 develops a density spike in the zone of
 influence of the SMBH. For  the
present considerations,  the \dkm  specifically consists of \pbhsn, 
  over what we find to be an astrophysically relevant mass range.
 
A more conservative case appeals to dynamical relaxation
 of the accreted \dkmn. This would 
occur after a  major merger,
 although adiabatic growth should eventually resume. In this case, 
$\rho \propto r^{-3/2}$
 for a black hole merging scenario \cite{Merritt}. 
 %for example of intermediate mass
 %black holes to form the SMBH \cite{Merritt}.
A 3/2 profile also results from \pbh scattering off
 of a  nuclear star cluster \cite{Gnedin}.

 For  the  model, we also  need the radius  $r_{spike}$
  where the density spike begins, within 
 the radius of gravitational  influence of the
 central supermassive black hole, defined by the
 competition between SMBH gravity  and the central stellsr velocity
 dispersion of M87.
 The spike radius can be defined as \cite{Gondolo} the radius
where the potential of the SMBH  falls to that of the inner galaxy,
as infered from the stellar velocity dispersion via virial arguments.
One then has  
 $r_{spike} =GM_{SMBH}/\sigma_\ast^2,$ where
 the black hole mass  in M87 is   \cite{eht}
 $M_{SMBH} =  6.5\pm 0.7\times 10^9 \rm \,  M_\odot$ and the
 central stellar velocity dispersion \cite{ Longobardi} is
 (with our estimated error from reported long slit absorption
 spectroscopy) $\sigma_\ast = 400\pm \rm 40 \, km/s.$  We thus take
 $r_{spike}=200 \rm \, pc$.

\subsection{Normalization of the spike}

In addition to the density profile, we also need to evaluate the 
absolute size or
 normalization of the spike. This can be
represented by the  value of the density 
 at $r_{spike}$, which we call $\rho_{spike}$.
This can then be  extrapolated  according to the assumed power law
to find $\rho_{horizon}$ as needed in \eq{siga} below.

We  assume a certain mass for the spike and relate this to $\rho_{spike}$.
 Integrating the power law, one finds
\beql{mas1}
M_{spike}\approx 4\pi\rho_{spike}r_{spike}^3\frac{1}{3-n}\,,
\eeql
where $n$ is the power in the profile.
 One notes that this, as opposed to the integral for $\Sigma$ below, is essentially
independent of $n$.
This is because  when $\rho$ is not too singular, $n<3$, 
 the geometric factor $r^2$ in the integral makes it depend
 essentially on the outer scale.

To proceed, we  parameterize the spike mass in terms of a constant $\eta$ as
\beql{param}
M_{spike}=\eta\times \nrnd{1}{9}M_\odot\,,
\eeql
so that $\eta =1$ would correspond to $M_{spike}/M_{SMBH}=0.16$ for M87..
The observed value of $\rho_{spike}$  can be inferred from  the central kinematics of M87 \cite{murphy} to approximately constrain $\eta \sim 0.1$.

We thus write
\beql{mas3}
\rho_{spike}=  \frac{(3-n)}{4\pi} \frac{1}{r_{spike}^3}
M_{spike}= \eta \times \nrnd{3}{1}(1-n/3) \rm M_\odot/pc^3\,.
\eeql
We record these values for $\rho_{spike}$  in the second column of the
 Table.  The third column gives the density scaled by the respective
 power laws to $r_{horizon}=0.001\,\rm pc$ and
 the fourth column the resulting \cd
according to \eq{siga} below.

\subsection{Surface density through the ``spike''}

The column or surface density $\Sigma$ is given by the integral of the
ordinary density $\rho$ along the flight path: $\Sigma= \int \rho(r) dr$.
Although the integral should in principle be taken over the entire
flight path, it will be dominated by the `spike' and so we estimate it 
simply from the beginning of the spike, $r_{spike}$ to the vicinity
of the SMBH, $r_{horizon}$. 
 We take the inner spike radius $r_{horizon}=0.001\,\rm pc$
 from general relativistic studies of the spike for Schwarzschild 
\cite{Sadeghian} or Kerr \cite{Ferrer} black holes.

Thus
\beql{sig}
\Sigma\approx \int_{r_{spike}}^{r_{horizon}}\rho(r)dr
\eeql

With $\rho$ represented by the  power law
$\rho \sim  r^{-n}$  we  introduce the rescaled radius
 $x=r/r_{horizon}$. Normalizing to the value at the horizon, we can
write 
\beql{rh}
\rho=\rho_{horizon}\frac{1}{x^n}\,.
\eeql
The value of $\rho_{horizon}$ can   be found by magnifying the ambient
\dkm density at the beginning of the spike
 by the factor $(r_{spike}/r_{horizon})^n$. It is this large factor which
can make the spikes significant in the calculation of a \cd.
We now carry out the integral \eq{sig} in the limit
  $ (r_{horizon}/r_{spike})<<1$.
\beal{siga}
\Sigma&=&r_{horizon}\times \rho_{horizon}
\int_1^{(r_{spike}/r_{horizon})}
\frac{1}{x^n} dx \\
\nonumber
&=&\bigl(\frac{1}{n-1}\bigr)\,r_{horizon}\times \rho_{horizon}(1-(r_{horizon}/{r_{spike})^{n-1}})\\
\nonumber
&\approx&\bigl(\frac{1}{n-1}\bigr)\, r_{horizon}\times \rho_{horizon}\, .
\eeal

The resulting values for 
$\Sigma$ are shown in the last column of the Table.

\begin{table}
\begin{center}
\begin{tabular}{|l|l|l|l|}
\hline
Model& $\rho_{spike}$ &$\rho_{horizon}$&$\Sigma$\\
\hline
\hline
$\rho\sim r^{-3/2}$&$\eta\times 15$&$\eta \times \nrnd{1}{9}$&
$\eta \times \nrnd{3}{6}$\\
\hline
&& &\\*
\hline
$\rho\sim r^{-7/3}$&$\eta\times 6.5 $ &$\eta \times \nrnd{2}{13}$&
$\eta \times \nrnd{1}{10}$\\
\hline
&&&\\
\hline
\end{tabular}
\end{center}
\caption{Parameters for two models of the \dkm spike at the SMBH of 
M87.
The beginning of the spike, $r_{spike}$,  has been
taken to be {\bf  $200 pc$} from the \bh
 and its inner radius $r_{horizon}$  near the SMBH's
horizon, namely $0.001 pc$. The parameter $\eta$ characterizes the 
 mass in the spike via \eq{param}.
 The units of the \dkm density $\rho$ are $\rm M_\odot/pc^3$
 and the units of the column or surface density $\Sigma$ 
are $\rm M_\odot/pc^2$.}
\label{tab}
\end{table}

\section{Application of  EHT \rsl of 10 \mua}
With these estimates for $\Sigma$, we can apply \eq{lim} using 
the  \rsl of 10\, \mua stated by the EHT collaboration \cite{eht}.
Inserting 10\, \mua on the lhs of \eq{lim} and requiring that the rhs
not exceed this, one finds,  with $\Sigma$ in units
of $\rm  M_\odot/pc^2 $
\beql{finds}
 \Sigma \times (\rm M/M_\odot) <\nrnd{2}{4}
\eeql 

The possibility is commonly  entertained   that  the \dkm  is composed
only partially of primordial black holes, characterized as a fraction
 $f_{bh}$ of the total \dkm mass. 

In the present case, we should then  use a \cd reduced by $f_{bh}$.
 Since the \dkm mass enters linearly everywhere, one could  effect
this, for example, by making the replacement $\eta \to \eta \times f_{bh} $
in the formulae.
Using  the inferred  value of  $\Sigma$ for the two models, one has the limits
\beal{lims}
M \times f_{pbh}
 \leq \frac{1}{\eta}\times \nrnd{7}{-3}\rm M_\odot~~~~~~~~~~~~~~&&\rho\sim r^{-3/2}\, model\\
\nonumber
M \times f_{pbh}
\leq \frac{1}{\eta}\times\nrnd{2}{-6}\rm M_\odot  ~~~~~~~~~~~~~~&&\rho\sim r^{-7/3}\, model
\eeal

 These limits should apply regardless of the
 possible clustering of primordial black holes.
 Clusters would not survive within the density spike around the
 SMBH. Poisson fluctuation-induced clustering has been shown
 to greatly reduce the number of predicted merger events
\cite{Jedamzik}, and allows the resurrection of a significant solar
 mass component  of PBHs that could account for the observed LIGO/VIRGO 
 \bh merger rates \cite{Clesse}. The astrophysical and primordial
 \bh interpretationa are currently indistinguishable \cite{franciolini}.
Our results show the possibiity for  strong  limits
on  compact objects  by a new  independent method. In particular the second
variant in \eq{lims} is of interest for  the asteroid/earth  mass range.
 This can be compared with the results of the
 OGLE \cite{OGLE} and Subaru \cite{SUBARU} microlensing collaborations.
  The M31 microlensing data, revised for finite lens size effects
 \cite{smyth},   sets a limit of around  1\% on the mass fraction 
 of PBHs contributing to dark matter in the mass range 
$10^{-9} -10^{-6} \rm M_\odot.$

\section{Comparisons}

\begin{figure}[h]
\includegraphics[width=\linewidth]{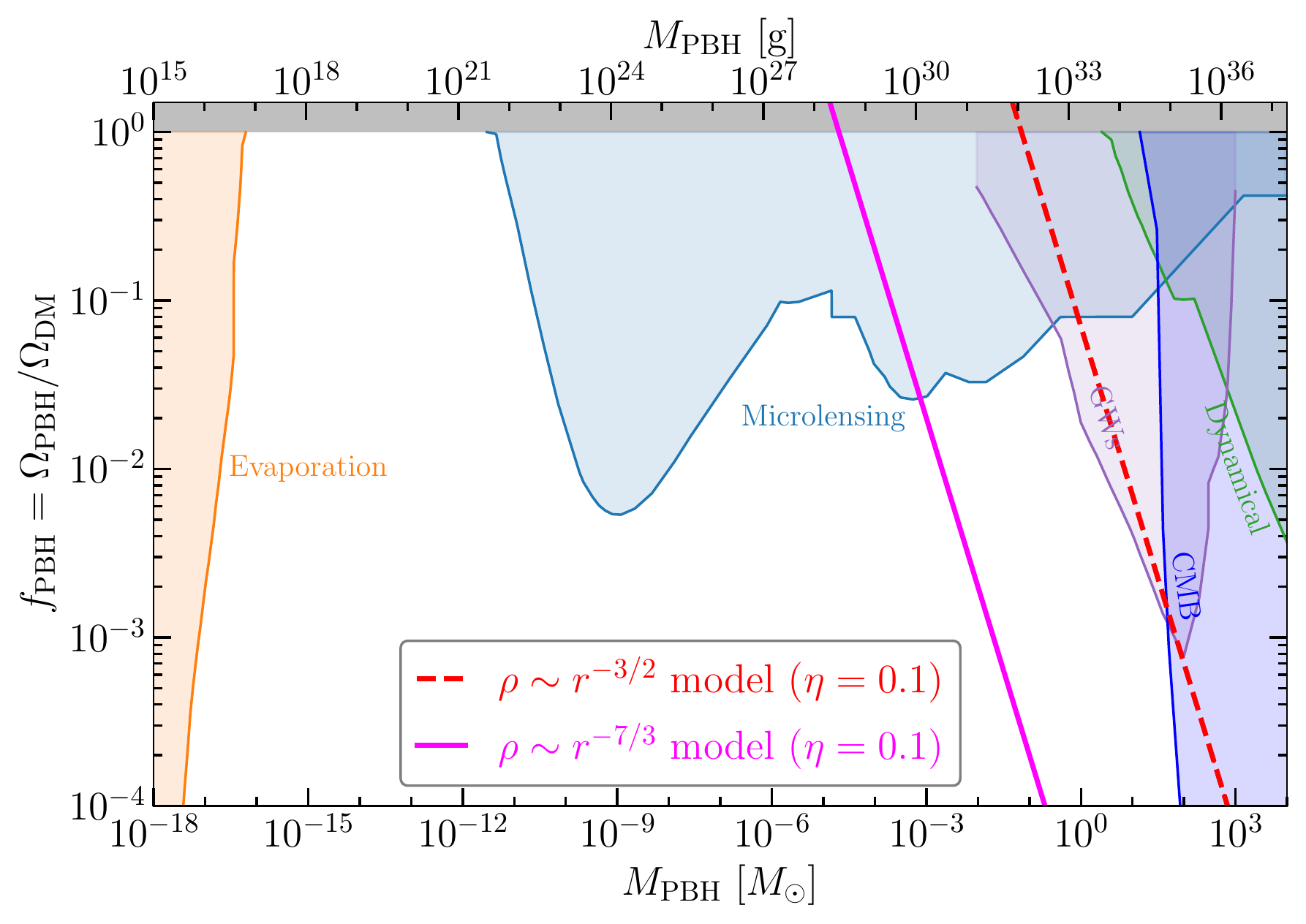}
\caption{Limits on \pbhs  exhibited in the $(M/M_\odot, f_{pbh})$ plane,
using \eq{lims}, for the $\rho\sim r^{-3/2}$ model ( dashed, right,  line)
and for the $\rho\sim r^{-7/3}$ model (solid, left, line)
 The respective  excluded regions are to the right of these lines. Other limits are quoted
from \cite{green} and \cite{bradkav}. Figure graciously provided by Guillem Domenech.   }
\label{exclu}
\end{figure}

A number of other methods have been used to limit the possibilities
for \pbhs. In Fig\,\ref{exclu} we show some representative  limits, in
 the $(M/M_\odot, f_{pbh})$ plane, as compared to ours
from \eq{lims}. We show  the case $\eta =0.1$. 
The dashed (right)  line is for the $\rho\sim r^{-3/2}$ model
  and the solid (left)  line is for the $\rho\sim r^{-7/3}$ model.
The excluded region, according to \eq{lims}, lies to the right and above
these lines.

Our arguments appear to exclude a considerable region of the  right-hand
 part of the $(M/M_\odot, f_{pbh})$ plane, in agreement with, and with the
$\rho\sim r^{-7/3}$ model, extending
previous exclusions. On the other hand, in the limiting case where
the \dkm  consists entirely of \pbhs or $f_{pbh}=1$, the method does not seem
to provide  limits as strong as those from microlensing. Improvement in the
input information, such as higher \rsl observations, would tend to move the 
exclusion lines to the left.
   
 The other exclusion limits shown  on the plot have been obtained from
 reference\,\cite{green} and from  reference\,\cite{bradkav}.

\section{Interpretation}

These interesting limits illustrate the potential power of the
 method. They
arise essentially  from the sharpness of the spike, with the associated large values
of $\rho_{horizon}$. The large difference
between the two variants in \eq{lims} arises from the sensitivity to the
steepness of the spike profile. Thus an improvement in the definition of our results would come
from a better understanding of the spike profile, either by
observational or theoretical arguments. Of course, better knowledge of
$M_{spike}$ or even better observational
\rsl would also be helpful.

As  stresed in \cite{sra} and \cite{if} the interpretation of a possible
positive signal would be a much more subtle matter that the simple 
setting of limits as we do here. That is, if a loss of, or a limit
 on,  observational \rsl is found, all other reasons for this, natural or
instrumental, have to be considered before attributing the effect to \pbhs.   
Further observations of other SMBH's with high \rsl would be of interest in
connection with our arguments. Note that in \eq{lim}  the \rsl enters
quadratically, and similarly  on the rhs of \eq{finds}, 
 so improved high \rsl will have a strong effect on the limits. 
  
  It is difficult to avoid the development
 of a dark matter spike around the SMBH in M87. It may be modulated by
 SMBH merging, but this would have most likely
 occurred within the first billion years after SMBH formation. Hence the effective value
 of accumulated DM  mass in the spike
 could be slightly reduced. We  argue   that 
 the adiabatic build-up of the central SMBH 
 in massive elliptical galaxies inevitably results
 in DM spike formation,  with intriguing consequences
 for the EHT observations of the supermassive black hole in M87.
 
 In the event that PBHs form all or most of the DM,
an interesting aspect of our method is that
  any pre-existing clustering of PBHs, such as that
 invoked to account for  consistency with the
 early MACHO experiments and with the LIGO/VIRGO event rates, 
  would not be preserved in the spike. With improved understanding of
spikes and further observations involving them, our new kind of limit
promises new, independent constraints on PBH masses
 as a DM contributor.

\section{Acknowledgement}

We thank Guillem Domenech for providing us with 
Figure\,\ref{exclu}.

\end{document}